\documentclass[12pt]{report}
\usepackage{graphicx}
\usepackage{indentfirst}
\usepackage{afterpage}

\def\gee{ \, \lower 1mm\hbox{$\,{\buildrel > \over{\scriptstyle\scriptstyle\sim} }\displaystyle \,$}}
\def\lee{ \, \lower 1mm\hbox{$\,{\buildrel < \over{\scriptstyle\scriptstyle\sim} }\displaystyle \,$}}
\def\|{\partial}

\def\varkappa {{\scriptstyle\partial}\! e}

\let\c=\centerline

\let\b=\baselineskip
\begin{document}
\begin{center}
\Large \textbf{Relativistic Binary Merging Rate in the Universe:25 years of confrontation}

V.M.Lipunov         $^{1}$ \normalsize

 {$^1$ - Sternberg Astronomical Institute, Moscow, 119899, Russia}\\
              lipunov@sai.msu.ru

   \date{Received May,  2004}
\end{center}

{Recently estimation of merger rate of double neutron stars from the observations
 of PSR J0737-3039 by \cite{Burgay} is discussed under real
astrophysical background.}

   \textsc{keywords} ~~~~ Stars: evolution -- Stars: neutron --  binaries: general
\normalsize

  %______________________________________________________________
\bigskip

\c{\bf{Introduction}}

Relativistic stars (neutron stars and black holes) merging can be discussed like "astrophysics" reaction of "elementary
particle" interaction. This merging is analogous to elementary physics processes in the world of elementary particles
(\cite{Lipunov1993}). There is no doubt, that there are the following processes in the Universe:

~~~~~~~~~~~~~~~~~~~~~~~~~~  NS + GWB

NS  +  NS  $\Longrightarrow$

~~~~~~~~~~~~~~~~~~~~~~~~~~ BH + GWB

The result depends on the mass of neutron star and Oppenheimer-Volkoff  limit.

NS  + BH  $\Longrightarrow$  BH + GWB

BH + BH  $\Longrightarrow$ BH + GWB,

where GWB is the Gravitational Wave Burst.

The "cross-section" or probability calculation of these processes in the Universe is of principal importance not only
for astrophysics, but, first, for fundamental physics, so as exactly these processes are accompanied by the most
powerful gravitational-wave emission. This emission has an impulse character, which can be detected by
gravitationally-waved antenna like LIGO.

The powerful of gravitation wave emission in these processes approximate to maximum possible value in nature (even if
we take into account the future theory of quantum gravitation (\cite{Lipunov1993}):

\begin{equation}
L_{max} = M_c^2/{R_g c} = E_{pl}/t_{pl}  = C^5/G = 10^{58} erg/s
\end{equation}

The detection of such processes possibility has been done by only 2 ways last 20 years. First one is to use our
understanding of binary stars evolution  processes and to use observed astronomical data in all wave-lengths. Second
one is based upon radio-astronomical data by radio-pulsars. Let's consider them.

\c{\bf{Two methods of merging rate estimations.}}

Both methods are based upon the observed data of Our Galaxy with the following generalization to the whole Universe.
But historically first one is called by "theoretical", and the second one is called by "observed". It's not right as a
matter of fact, but let's use this terminology.

The possibility of the processes or the cross-section can be characterized in the terms of "merging rate", normalized
to the galaxy like Our one. Practically, normalization on $10^{11} M_{\odot}$ of luminous barionic matter is suggested.

"Theoretical" estimation is always attached to the following
chain:

\textit{ - Merging Rate is equal to  Star Formation Rate in galaxy (Solpiter Function);}

\textit{- the part of binary stars, that can form the relativistic star (the distribution function by the relation of
masses of binary components)};

\textit{- the part of the stars, that survived after the first Super-Nova explosion (it's strongly depends on the
anisotrophy of the collapse or so called kick-velocity)};

\textit{- the part of neutron stars after the second explosion} and

\textit{- the part of double relativistic stars, that can be merged at the Hubble time.}

The most weak link in this chain is our unknowledge of possibility of collapse anisotrophy. But the "theoretical"
method, that was realized in the most completely realization (see monograph "Scenario Machine", Lipunov et al.,1996,
Tutukov et al.), suggests the obligatory calibration of unknown parameters by the observed data in all wave-length
(from radio to x-ray). So, if the mean output velocity would be too large, all massive x-ray stars like x-ray-pulsars
would be disappear from the sky (Cen X-3, Vela X-1, etc., the number is about 50), so as at large anisotrophy of the
collapse the binary stars will be too quickly broken. On the contrary there will be too much of such systems at small
anisotrophy, and there will be contradiction with observed number of binary radio-pulsars. The first method
(\cite{Lipunov1987}) gave the estimation in $10^{-4}$ on the galaxy like Our one. (see Fig. 1)

 \begin{figure}
   \centering
   \includegraphics[angle=0,scale=0.6]{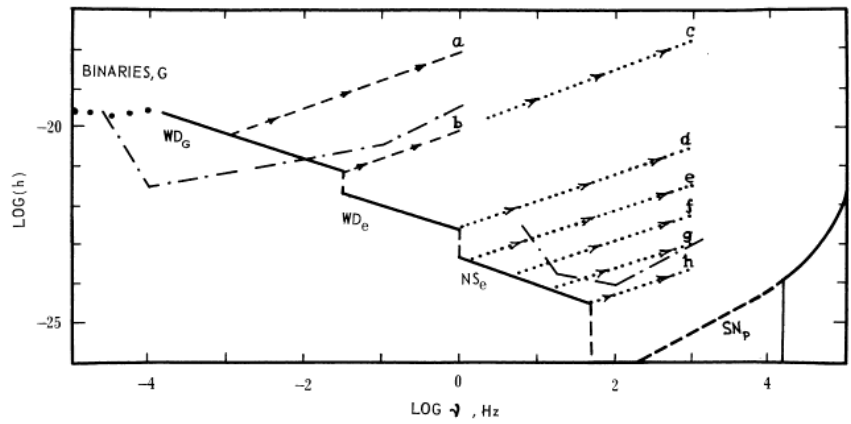}
      \caption{Gravitational Wave Spectra from astrophysical sources (\cite{Lipunov1987}).
NS merging rate ($year^{-1}$) for distances less than 20 Mpc (line e). It corresponds to Merging Rate in $1/10^4$ yrs
per $10^11$ solar Mass.
              }
   \end{figure}

Second "observed" method was first used by \cite{Phinney}. It based on observed  parameters of binary radio-pulsars,
that can be merged at the Hubble time. At the 1991 there was only one such pulsars, and the estimation was  $10^{-6}
year^{-1}$ in our Galaxy ($10^{10} M\odot$).

Just this not right estimation, to our mind, served to begin the
building the gravity interferometer LIGO.

The main problem of observed method is not in the fact, that there was used only one observed example for statistical
estimation , and is not in the fact, that interpretation of observations always was difficult from the selection
effects (uncertainty of the distance, collimation angle, life-time, horizon of sensitivity that is much more smaller,
than Galaxy; we see less than 1\% of all radio-pulsars). The main problem is in the interest to the process of neutron
stars merging, no radio-pulsars. Simple analyse shows, that neutron star passes not less than 6 physically different
states during its rotating evolution. The phenomena of radio-pulsars is very specific from these states, and in not
always able to be observed (\cite{Lipunov1991})%, 1991, "Astrophysics of Neutron Stars").

There are the change of the probability estimation of the process of neutron stars merging during the last 25 years in
the Table and Graphic (Fig.2). One can compare them. I assert, that most adequate to modern standard of interpretation
of binary and relativistic stars evolution "theoretical" estimation didn't change during the last 17 years and
beginning from the 1987 year always gave the value  $10^{-4 \pm 0.5} / year $ in the levels of reduced precision.

This estimation corresponds to one merging per minute in whole Universe and to 1 event per year at the gravity-waved
detector with $10^{-21}$ sensitivity.

\begin{table}[ht]
\begin{center}
\caption{"Theoretical" estimations of Neutron Stars Merging normalized to $10^{11} M_\odot$ .}
\smallskip \begin{tabular}{cc}      \hline
Author & Estimation\\
\hline
\cite{Clark}                       &  $1/10^4  - 1/10^6$ \\
 \cite{Lipunov1987}                               &  $      1/10^4$ \\
\cite{Hills}                       &   $    1/10^4$ \\
\cite{Tutukov}                 &   $ 1/10^4 $\\
\cite{Lipunov1995}                                &   $ <  3/10^4 $\\
\cite{Portegies1996}           &   $     3/10^5 $\\
\cite{Lipunov1996}                                &   $3/10^4 - 3/10^5$ \\
\cite{Portegies1998}       &   $1/10^4 - 3/10^5 $\\
\cite{Bethe} &                        $      1/10^4$ \\
\hline
\end{tabular}
\bigskip
 \caption{"Observed" estimations of Neutron Stars Merging Rate.}
\smallskip \begin{tabular}
{cc}      \hline Author & Estimation\\  \hline
\cite{Phinney}   &                                     $1/10^6$ \\
\cite{Narayan}  &                          $      1/10^6 $\\
\cite{Curran} &                        $       3/10^6 $\\
\cite{Van den Heuvel} &               $        8/10^6 $\\
\cite{Bailes}   &                       $ <$  $1/10^5 $\\
\cite{Burgay}                           & $1/10^4$\\
\hline
\end{tabular}
\end{center}
\end{table}

\begin{figure}
   \centering
   \includegraphics[angle=270,scale=0.6]{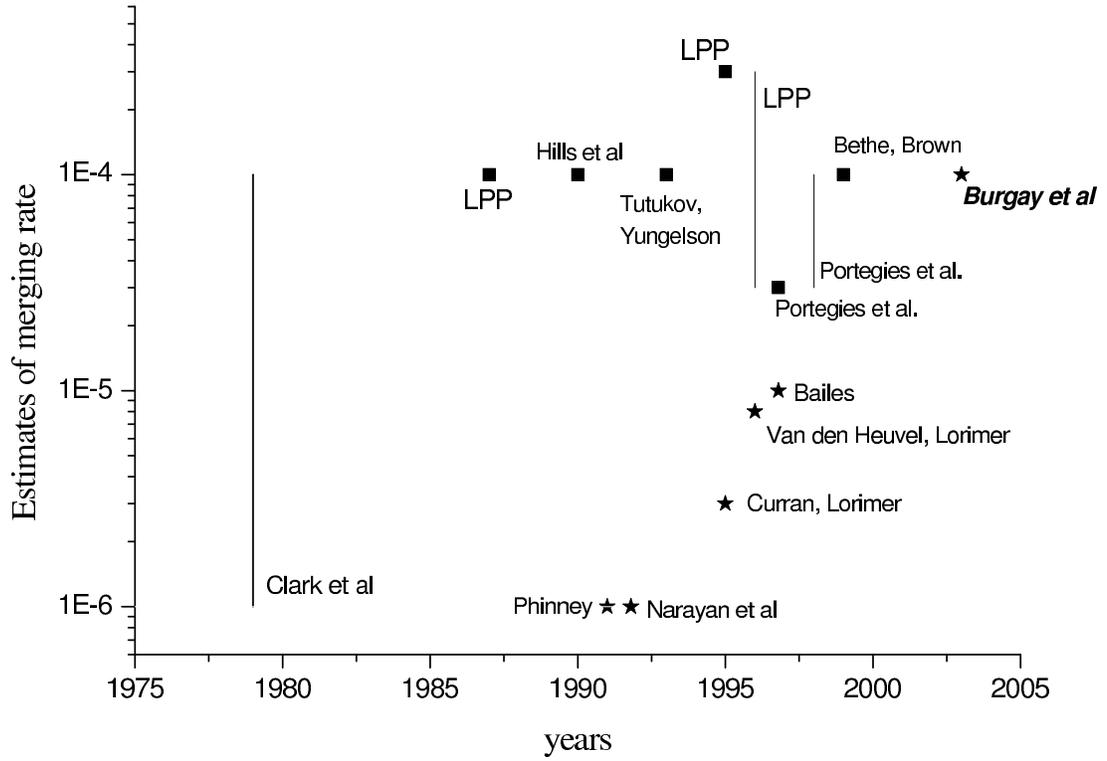}
      \caption{Merging Rate estimation by different authors. Squares are the "theoretical" method, "stars" are the
      observational one.}
%\label{Fig2}
   \end{figure}

 \begin{figure}
   \centering
   \includegraphics[scale=0.5]{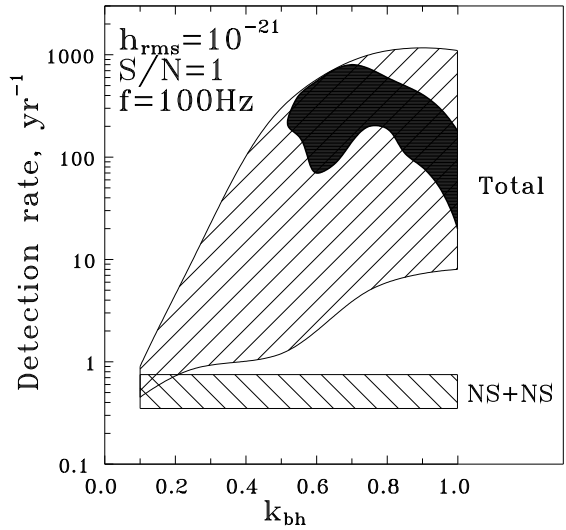}
      \caption{Predicted Detector rates for \textbf{ Neutron Stars } (horizontal branch) and \textbf{Neutron Stars - Black Holes}
        and \textbf{BH + BH - dark area} (\cite{Lipunov1997b}). $K_{bh}$ is the part of pre-supernova mass which collapsed into the Black Hole.}
   \end{figure}

More difficult problem is to estimate the frequency of the reaction with black hole participation. Our understanding of
the evolution is essentially worse here. Nevertheless, \cite{Lipunov1997b},\cite{Lipunov1997c} could get round the
theoretical uncertainty, using simple observed limits. They are the following: there is no any radiopulsar with black
hole on the sky (this is up limit) and there is at least several black holes in the binary with massive optical stars
(for ex., Cyg X-1) in Our Galaxy. As it was shown in  \cite{Lipunov1997b},\cite{Lipunov1997c}, it is more possible to
register the gravity-waved impulse from the black-holes merging:

\smallskip

BH  +   NS $\Rightarrow$ BH  + GWB

 BH  +   BH $\Rightarrow$   BH + GWB               $10^{-5}/year/galaxy$

\smallskip

So as the mean black hole mass can be in 8-10 times more, the frequently of these processes at the detector can be
essentially more, than from the neutron stars merging (see Fig.3).

Recently, \cite{LipunovPanchenko} proved that preliminary possibility of last two processes can be increased up to 5-7
times.

$$
$$

\c{\bf{Conclusions}}

1) So called "theoretical" estimations give us the merging rate for Neutron Stars in $10^{-4 \pm 0.5}$ from
\cite{Lipunov1987}. I accent, that "theoretical" estimation isn't simple average between the results of different
authors. All articles, that gives the merging rate for NS+NS less than $1/(3 * 10^4) year^{-1}$ are wrong, becouse in
this case all binary radiopulsars and binary X-ray pulsars disappear from the sky \cite{Lipunov1997a}. There are
another points of view (for ex., \cite{Kalogera}), I can't agree with them.
 One must accentuate, that the most full and correct model of binary stars evolution is the
"Scenario Machine", that takes into account the evolution of magnetized neutron star (see for details
\cite{Lipunov1996})

2) The "observed" estimations,
that used radio-pulsars data, were always burdened by selection
effects.

3) The gravitation impulses from the merging with black holes participation must be the first events on the
interferometers like LIGO (\cite{Lipunov1997b},\cite{Lipunov1997c}).

%Figures
%
%fig1 -  ~lipunov/merging/fig1.eps Gravitational Wave Spectra from astrophysical sources (lipunov1987).
%NS merging rate 1/year for distance less than 20 Mpc (line -e). This corresponds to Merging Rate $1/10^4$ yrs per
%$10^11$ solar Mass.

% fig2 - ~lipunov/merging/estimation.ps eto grafik, kotoriy mis stroili s toboy
%"The estiamtoin Mergin Rates by different authors. Kvadratiki - "theoretical" method, "zvesdochki" - "observational" -
%method. $K_bh$ - is the part of presupernova mass wich collapsed into the Black Hole.
%file fig3  ~lipunov/merging/golova.eps

%Podpis Pridction od the Detection Rates for LIGO-type detector with sensitivity $10^ -21$. Horisontal brench - NS-NS -
%merging, Black ("Dinozaurus had") - BH+BH and BH+NS - merging. (LIPUNOV ET.AL.,, 1997)
%\end{document}

%fig2 -  "Theoretical" (квадратики и вертикальные бары) and "Observational" (звездочки) estimations of the neutron
%stars merging rate depend on the year of the publication.

%fig3 - Predicted Detector rates for Neutron Stars (horizontal branch) and Neutron Stars - Black Holes and BH+BH - dark
%area (LIPUNOV ET.AL.,, 1997b).
\end{document}